\documentclass[12pt]{iopart}
\usepackage{graphicx}
\usepackage{color}
\usepackage{cite}
\usepackage{iopams}

\newcommand{\ket}[1]{\mbox{$|#1\rangle$}}
\newcommand{\bra}[1]{\mbox{$\langle#1|$}}
\newcommand{\braket}[2]{\mbox{$\langle#1|#2\rangle$}}
\newcommand{\set}[1]{\{#1\}}
\newcommand{\outerp}[2]{\ket{#1}\bra{#2}}

\begin{document}

\title{Coherent transport over an explosive percolation lattice}

\author{\.{I}. Yal\c{c}{\i}nkaya and Z. Gedik}
\address{Faculty of Engineering and Natural Sciences, Sabanc{\i} University, Tuzla, 34956, \.{I}stanbul, Turkey}
\ead{iyalcinkaya@sabanciuniv.edu}
\vspace{10pt}
\begin{indented}
\item[]January 2016
\end{indented}

\begin{abstract}
We investigate coherent transport over a finite square lattice in which the growth of bond percolation clusters are subjected to an Achlioptas type selection process, i.e., whether a bond will be placed or not depends on the sizes of clusters it may potentially connect. Different than the standard percolation where the growth of discrete clusters are completely random, clusters in this case grow in correlation with one another. We show that certain values of correlation strength, if chosen in a way to suppress the growth of the largest cluster which actually results in an explosive growth later on, may lead to more efficient transports than in the case of standard percolation, satisfied that certain fraction of total possible bonds are present in the lattice. In this case transport efficiency increases as a power function of bond fraction in the vicinity of where effective transport begins. It turns out that the higher correlation strengths may also reduce the efficiency as well. We also compare our results with those of the incoherent transport and examine the average spreading of eigenstates for different bond fractions. In this way, we demonstrate that structural differences of discrete clusters due to different correlations result in different localization properties.
\end{abstract}


\vspace{2pc}

%
%
%

\section{\label{sec:int}Introduction}

Coherent transport over complex networks has been a topic of much interest in the recent years \cite{mulken2011}. Such processes are often related with the dynamics of excitations over networks modelled by quantum walks and studied for both variants, namely the discrete- \cite{ambainis2001,konno,bach,yamasaki,kwek,chandrashekar2014quantum,asboth,gonulol} and the continuous-time \cite{mulken2005slow,mulken2005spacetime,mulken2006efficiency,mulken2006coherent,krapivsky,mulken2007survival,blumen,volta} quantum walks. Although the original proposals \cite{aharonov,farhi} of either models are mainly aimed to outperform their classical counterparts in terms of spreading rates, it has been shown later that quantum walks are useful tools also for developing new quantum algorithms \cite{ambainis}, quantum simulations \cite{schreiber,crespi,ghosh,kitagawa,genske} and the universal quantum computation \cite{childs,lovett}. In the context of coherent transport, they provide simple models to describe quite significant physical phenomena such as the excitonic energy transfer through the photosynthetic light-harvesting complexes \cite{mohseni} or breakdown of an electron system driven by strong electric fields \cite{oka}. 

It is possible to introduce a disorder to a network using the standard bond percolation model in which the bonds between sites are either present or missing with some probability $p$ \cite{grimmett,saberi}. A group of connected sites is called a cluster and its size is defined by the number of these sites in total. For an infinite network, if $p$ is smaller than some critical value $p_c$ (the percolation threshold), there exist only small clusters which remain small even if we increase the network size. However, just after $p=p_c$, small clusters merge to form a large cluster covering the whole network, namely the wrapping cluster, comparable to the network in size. This kind of disorder constitutes a source of decoherence for quantum walks and appears in two variants: the static and dynamic percolations. In the former, the network configuration does not change during propagation, whereas in the latter, connections between sites do alter in time. Both variants have extensively been studied so far in the context of transport and spreading properties for discrete- and continuous-time quantum walks \cite{romanelli,annabestani,leung,schijven,kollar,mulken2007quantum,anishchenko,darazs2013,darazs2014,elster,stefanak}.

Achlioptas \etal proposed a model for network construction by slightly changing the standard percolation, which leads to a very fast growth of the wrapping cluster \cite{achlioptas}. According to this model, two bond candidates are randomly selected each time a new bond is intended to be added. Then, the one minimizing the product of the cluster sizes it merges is placed as a new bond and the other one is omitted. This simple rule suppresses the formation of the wrapping cluster but eventually results in the abrupt (or so-called \textit{explosive}) growth of it after $p=p_c$. In contrast with the standard percolation, the network configuration for a given $p$ depends on the total previous occupation history. In this sense, the disorder due to explosive percolation cannot be considered as a fully random process but rather a correlated one.

It is well known that the efficiency of coherent transport can be increased by exploiting environmental effects \cite{mohseni,viciani,novo,biggerstaff,chandrashekar2014noise}. One of the methods in achieving this goal, for example, utilizes the interplay between the coherent dynamics and the disorder due to the topology of the network where transport takes place \cite{stefanak,asboth}. In this article, we use a similar, but distinctively different approach where the structure of the clusters contributing to the transport are affected by the cluster correlations during the growth of the network. We look for correlation strengths that yield more efficient transports than that of coherent dynamics alone. With this motivation, we examine the transport of an excitation along a certain direction over a square lattice where the bond configuration is determined by explosive percolation statically. We therefore introduce the sites on the left edge of the lattice as sources and the ones on the right as sinks, where an excitation is created and absorbed, respectively \cite{anishchenko}. In this way, we monitor the survival probability over the lattice in the long time limit to find out the transport efficiency after starting with an initial state localized on the source sites. In modelling the transport, we use the continuous-time quantum walk which is also closely related with the tight-binding models in solid state physics. We compare the transport efficiencies with increasing bond fraction for standard and explosive percolation models to find out whether any model has supremacy over the other. 

This article is organized as follows. In \sref{sec:met}, we overview coherent and incoherent transport over dissipative lattices along with the description of percolation models to be used. In \sref{ssec:treff}, we compare the efficiencies of transport models in case of standard and explosive percolation. In \sref{ssec:boswc}, we investigate the spreading of eigenstates. In conclusion, we summarize our results.

\section{\label{sec:met}Methods}

\subsection{\label{ssec:cohtr}The coherent and incoherent transport}

We consider a square lattice of $N$ sites, which we will denote by $L=\sqrt{N}$, as the environment where the transport process takes place. The sites are labeled by positive integers $\cal{N} =$ $\{1,2,\cdots,N\}$. The information about the existence of bonds in between is held by the Laplacian matrix $L$ where the non-diagonal elements $L_{ij}$ are $-1$ if site $i$ and site $j$ are connected and zero otherwise. The diagonal elements $L_{ii}$ hold the total number of bonds that belong to the site $i$. Thus, $L$ is a positive semi-definite matrix, i.e., its eigenvalues are non-negative. An excitation localized on any site $i$ is interpreted as being in the state $\ket{i}$ and these states form an orthonormal and complete set over all sites, i.e., $\braket{k}{j}=\delta_{kj}$ and $\sum_{i\in\mathcal{N}}\outerp{i}{i}=I$. A coherent (incoherent) transport is modelled by continuous-time quantum (random) walk which is described by the Hamiltonian (transfer matrix) $H_0=L$ ($T_0=-L$) \cite{mulken2011,farhi}. Here, we assume that transition rates are identical and equal to $1$ for all sites. The transition probability from the initial state $\ket{\psi_j}$ at $t=0$ to the state $\ket{\psi_k}$ is  $\pi_{kj}(t)=|\bra{\psi_k} \exp{(-iH_0 t)} \ket{\psi_j}|^2$ for the coherent and $p_{kj}(t)=\bra{\psi_k}\exp(T_0t)\ket{\psi_j}$ for the incoherent transport, where we assumed $\hbar=1$.

\begin{figure}[!t]
\centering
\includegraphics[scale=1.7]{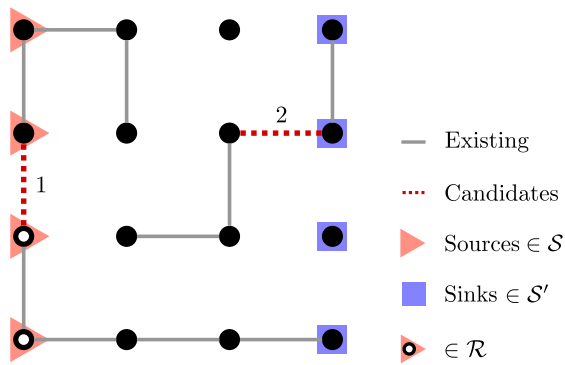}
\caption{An example of a bond percolation lattice for $L=4$. Left (right) edge contains source (sink) sites where an excitation is created (absorbed). Dashed lines represent two randomly selected bond candidates labeled by $1$ and $2$ with corresponding weights $w_1=4\times 5=20$ and $w_2=2\times 3=6$. According to the best-of-two rule, bond $2$ will be selected and bond $1$ discarded for $w_2 < w_1$. A wrapping cluster is defined to be the one connecting the left edge to the right edge. For this example, it lies along the bottom edge of the lattice and occupies two source sites. The set $\mathcal{R}$ contains the source sites belong to the wrapping cluster for the given lattice realization.}
\label{fig:perclat}
\end{figure}

Once an excitation covers the lattice from one side to another, we understand that it did get transported over the lattice. In order to keep track of this process, we define the sites at the left (right) edge of the lattice as the sources (sinks) as in \fref{fig:perclat} \cite{anishchenko}. We will denote the set of all source and sink sites by $\mathcal{S}$ and $\cal{S}'$, respectively. Sources are the only sites where an excitation can initially be localized and the sinks are abstract representations of absorption or trapping processes. Thus, a `leak' taking place on the right edge implies that an excitation, originally localized on the left edge, has been transported along the lattice. This process can be introduced by a projection operator $\Gamma=\sum_{k\in\cal{S}'}\outerp{k}{k}$ perturbing the Hamiltonian $H=H_0-i\Gamma$ or the transfer matrix $T=T_0-\Gamma$ where we choose the leaking rates to be the same and equal to $1$ for all sink sites. In the limit $t\rightarrow\infty$ and for the initial state $\ket{\psi_j}$, the total probabilities of finding the excitation on the lattice, namely the survival probabilities for coherent and incoherent transports are given as (see Appendix),
\begin{equation}
\Pi_j=\sum_l|\braket{\psi_j}{\Phi_l^R}|^2,\hspace{4mm}
P_j=\sum_{k\in\mathcal{N}}\sum_l\braket{k}{\phi_l^0}\braket{\phi_l^0}{\psi_j},
\label{eq:limsurprob}
\end{equation}

\noindent
where $\ket{\Phi_l^R}$ and $\ket{\phi_l^0}$ are the eigenstates of $H$ and $T$ with real and zero eigenvalues, respectively. The initial state $\ket{\psi_j}$ may involve sites only from the set $\mathcal{S}$. In this article, we will denote the complement of survival probabilities by $\mu$  which can equally be interpreted as the transport efficiency \cite{stefanak} or the percolation probability \cite{chandrashekar2014quantum}. By calculating $\mu$, we will monitor how much information may escape from the lattice.

\subsection{\label{ssec:expperc}The explosive percolation}

The standard bond percolation is implemented on a square lattice by first removing all bonds between the sites and then, randomly adding them one after another. This process results in random growth of discrete clusters. For infinite lattices, the opposite borders get connected to each other through one large wrapping cluster after reaching a critical fraction of bonds $p_c=0.5$ \cite{essam} called the percolation threshold. Here, the bond fraction $p$ is defined as the ratio of the number of bonds $n$ present in lattice to the number of total possible bonds, $p=n(2L^2-L)^{-1}$. 

In explosive percolation, a similar implementation procedure is followed with slight modification. Now, in order to add a bond, $m$ random bond candidates are chosen and a weight is assigned to each of them equal to the product of cluster sizes they may potentially merge. Then, the bond with smallest weight is occupied and the others are discarded (see \fref{fig:perclat}). In case a bond connects two sites within the same cluster, the corresponding weight becomes the square of the cluster size. We will see later that this complementary rule has drastic effects on the results we obtain. This selection rule here, called the \textit{best-of-m product rule} \cite{andrade,costa}, systematically suppresses the merging of small discrete clusters and consequently avoids the formation of a giant cluster up to some percolation threshold $p_c$ dependent on $m$ \cite{radicchi2010explosive,ziff2010scaling}. Once the threshold is exceeded, finite discrete clusters start joining each other much faster than in standard percolation ($m=1$) and finally, this results in an explosive behavior in the growth of the largest cluster. In particular, $m=2$ corresponds to the Achlioptas \etal model \cite{achlioptas}.

For $m>1$, discrete clusters cannot grow in a completely random manner as in standard percolation case. The shape and size of a given cluster becomes somewhat correlated with those of other clusters during the growth process. In this context, we interpret $m$ as the correlation strength since it specifies the number of discrete clusters taken into account while deciding to add a new bond. We examine the behavior of transport processes on a square lattice formed for different $m$ values.

Lastly, we note that a transport process can only take place after a wrapping cluster is formed. In infinite lattices, this happens just after $p=p_c$. In our case of finite lattices, however, there is still a chance of having no wrapping clusters when $p>p_c$ and, indeed, having one for $p<p_c$. Consequently, the average efficiency of the transport inevitably gets affected by these finite-size effects.

\section{\label{sec:numres}Numerical results}

We numerically determined all eigenvalues and corresponding eigenvectors of $H$ and $T$ for each lattice realization to calculate $\Pi_j$ and $P_j$ in \eref{eq:limsurprob}. It is clear that these quantities depend sensitively on the total number of real and zero eigenvalues. For this reason, we carefully compared numerical values with the exact ones for different lattice sizes and concluded that $L=7$ is the optimal one for our calculations where there is one-to-one correspondence between the exact and numerical results provided that numerical values smaller than $1.0\times 10^{-14}$ are set to zero. In order to obtain notable quantities for the above mentioned static percolation models, we averaged our results over $4\times 10^3$ lattice realizations for each $p$. From now on, $\langle\cdots\rangle$ will be used to indicate the ensemble averaged quantities we have obtained. In the figures, the standard deviation of the mean of each data point gets smaller than the thickness of the data line as the sample size grows larger than 10. For this reason, we do not include error bars to maintain clarity.

\subsection{\label{ssec:treff}Transport efficiency}

\begin{figure*}[!htbp]
\fontsize{10}{10}\selectfont
\begin{tabular}{ll}
\input{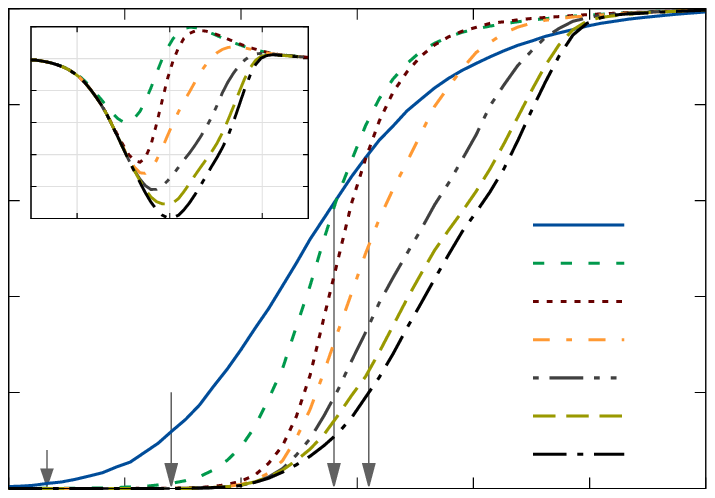}\hspace{5mm}&
\input{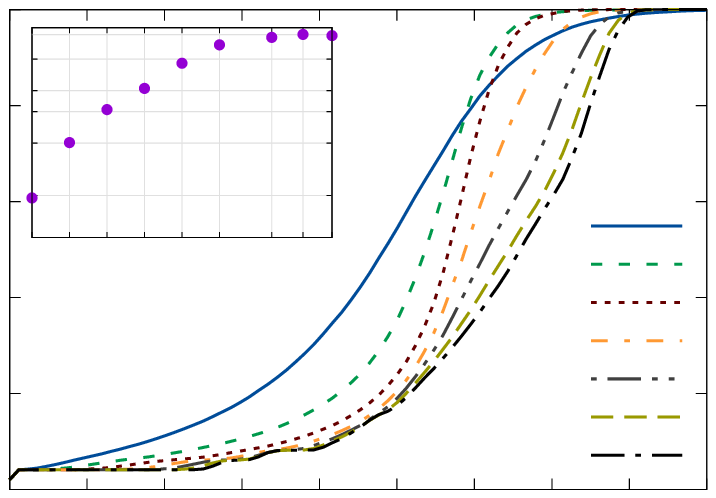}\\
\input{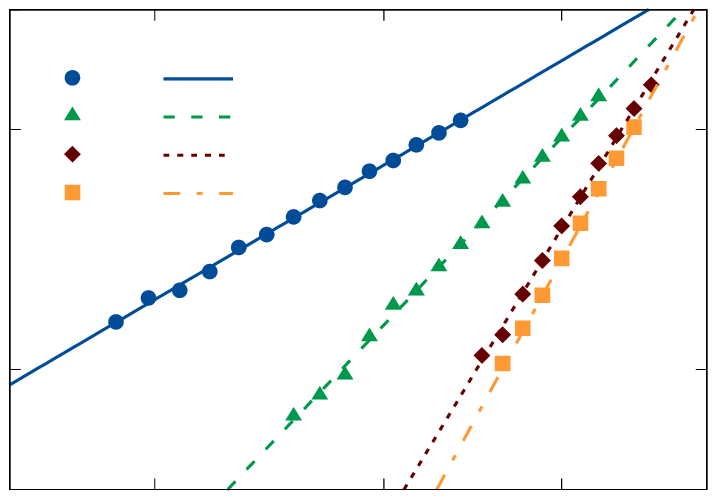}\hspace{5mm}&
\input{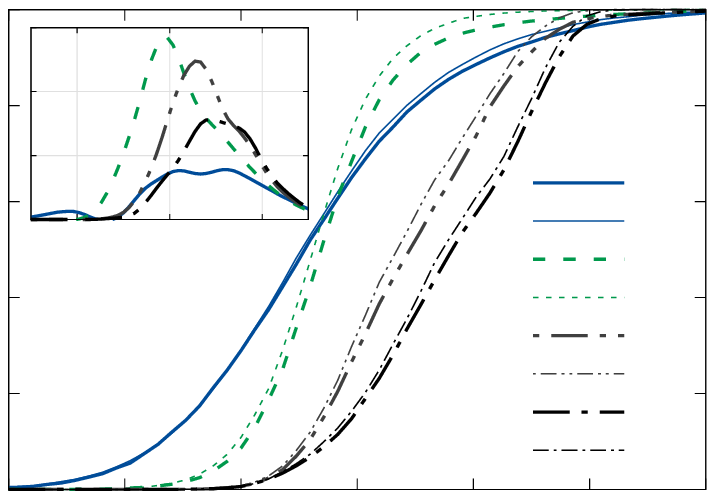}\\
\end{tabular}
\caption{(a) The average transport efficiency of coherent transport $\langle \mu_c^m \rangle = 1-\langle \Pi_\mathcal{S}^{m} \rangle$ vs. bond fraction $p$ for different correlation strengths $m$ after starting with the initial state $\ket{\psi_\mathcal{S}}$. The standard percolation and Achlioptas \etal models \cite{achlioptas} correspond to $m=1$ and $m=2$, respectively. The inset represents the comparison of $m=1$ and $m>1$, namely $\Delta\langle\mu_c^{m}\rangle=\langle\mu_c^{1}\rangle-\langle\mu_c^{m>1}\rangle$. (b) The average ratio of number of sites in the largest cluster over total number of sites $\langle \zeta^m \rangle$. The inset represents the average bond fraction of having a wrapping cluster $\langle p_w^m \rangle$ for different $m$ values, which are also represented in \tref{tab:exceeding}. The $m$ axis is drawn in the logarithmic scale. (c) The initial growth rate of $\langle \mu_{c}^m \rangle$ which fits a power function $\langle\mu_{c}^m\rangle \sim p^{k}$ for $m\lesssim 8$. The axes are drawn in logarithmic scale. (d) Comparison of the transport efficiencies of coherent and incoherent transports where the inset shows the differences.}
\label{fig:transeff}
\normalsize
\end{figure*}

\begin{table}
\caption{\label{tab:exceeding}Numerical values of some important parameters related to $m$.}
\begin{indented}
\item[]\begin{tabular}{ccccccc}
\br
$~~m~~$&$~~p_a^m~~$&$~p_b^m~$&$\langle\mu_{c}(p_b^m)\rangle$ & $~k^m~$ &
$~\langle p_w^m \rangle~$\\
\mr
1  & 0.33 & \textrm{n/a} & \textrm{n/a}   & 9.0          & 0.49\\
2  & 0.44 & 0.58         & 0.61           & 16           & 0.54\\
4  & 0.50 & 0.61         & 0.69           & 25           & 0.57\\
8  & 0.51 & 0.69 		     & 0.87           & 28           & 0.59\\
16 & 0.51 & 0.76 		     & 0.94           & \textrm{n/a} & 0.62\\
32 & 0.51 & 0.80  		   & 0.96           & \textrm{n/a} & 0.63\\
84 & 0.51 & 0.81 		     & 0.97           & \textrm{n/a} & 0.64\\
\br
\end{tabular}
\end{indented}
\end{table}
\normalsize

We choose an initial state which is equiprobably distributed over the source sites as $\ket{\psi_j}\equiv\ket{\Psi_\mathcal{S}}=\kappa\sum_{i\in\mathcal{S}}\ket{i}$ where $\kappa=L^{-1/2}$ for the coherent transport and $\kappa=L^{-1}$ for the incoherent transport. This state represents our lack of knowledge about the exact position of the excitation at $t=0$. In the limit $t\rightarrow\infty$, we define the average transport efficiencies for a given bond fraction and correlation strength $m$ as
\begin{equation}
\langle \mu_c^m \rangle \equiv 1-\langle \Pi_\mathcal{S}^{m} \rangle,\hspace{4mm}
\langle \mu_i^m \rangle \equiv 1-\langle  P_\mathcal{S}^{m} \rangle,
\end{equation}
for coherent and incoherent transports, respectively. These quantities are the average probabilities for the excitation to be absorbed in the limit $t\rightarrow \infty$. We will use the superscript $m$ for labeling purposes through the rest of this article. Let us also define $p_a^m$ here as the minimum bond percentage which satisfies $\langle \mu_{c}^m\rangle \geqslant 0.01$ \cite{chandrashekar2014quantum} in order to determine the effective starting point of the coherent transport. 

In \fref{fig:transeff}(a) the change in $\langle \mu_c^m \rangle$ is plotted with respect to $p$. The differences $\Delta\langle\mu_c^{m}\rangle \equiv\langle\mu_c^{1}\rangle - \langle \mu_c^{m>1} \rangle$ between the efficiency of the case $m=1$ and the efficiencies of the cases with larger $m$ are given in its inset. We see that when $m>1$, we obtain transports with partially higher efficiencies than that of the $m=1$ once $p$ exceeds certain values denoted by $p_b^m$. Additionally, $p_a^m$ increases and gets fixed for $m\geqslant 8$ within the given accuracy in T \tref{tab:exceeding}. The case $m=2$ starts to overcome $m=1$ at $p_b^2=0.58$ and the maximum peak occurs at $p=0.64$ where $\Delta\langle \mu_c^m\rangle\approx 0.1$. Actually, this is the extreme case among the others, i.e., a two-by-two correlation between discrete clusters contributes the most to the transport efficiency here unlike the rest of the cases. For increasing values of $m$ following $m=2$, the positive peak of $\Delta\langle\mu_c^m\rangle$ decreases in contrast with its increasing negative peak and $p_b^m$ shifts towards higher bond fractions. When $m=84$, there is almost no contribution to the transport efficiency: $\langle\mu_c^{84}\rangle$ can barely exceed the $\langle\mu_c^1\rangle$ after $p_b^{84}=0.81$. It is therefore evident that higher correlation strengths are increasingly inhibiting the transport process. 

In \fref{fig:transeff}(b), size of the largest cluster $\langle \zeta^m \rangle$ with respect to $p$ is given. The inset shows the average bond fractions $\langle p_w^m \rangle$ where a wrapping cluster is formed. When $p<0.5$, although $\langle\zeta^m\rangle$ reduces with increasing $m$, it tends to remain almost the same for $m\gtrsim 8$. This result exhibits a correlation with the behavior of $p_a^m$ in \fref{fig:transeff}(a) despite the saturating increase in $\langle p_w^m \rangle$: The size of the largest cluster may have a direct effect on the bond fraction where coherent transport effectively starts. When $p>0.5$, the $\langle \zeta^m \rangle$ decreases for higher $m$ values which is very similar with the behavior of $\langle \mu_c^m \rangle$. Also, the $p_b^m$ appear to be very close to the bond fractions where $\langle \zeta^{m>1} \rangle \approx \langle \zeta^1 \rangle$. Therefore, the comparison of \fref{fig:transeff}(a) and (b) strongly suggests that independent of $m$, the transport efficiency is determined by the size of the largest cluster at any $p$.

Our choice of $m=84$ as an upper limit of correlation strength for this article is intentional. There can be maximum $84$ bonds in the lattice $L=7$ in total, and hence, all discrete clusters most probably get correlated with each other as the lattice gets filled with bonds. We have repeated our calculations for $m>84$ and obtained quite similar results with the $m=84$ case. We also note that $\left < p_w^m \right >$ is approximately saturated after $m=84$. Therefore, it can be considered as an upper limit and the transport properties do not change thereafter.

When we again look at \fref{fig:transeff}(a), we see that the behavior of each $\langle\mu_c^m\rangle$ can be examined in three successive regions: (i) An initial growth with increasing rate in the vicinity of $p_a^m$, (ii) an approximately linear behavior and (iii) saturation. In region (i), transport efficiency fits a power function $\langle\mu_c^m\rangle \sim p^k$ as shown in \fref{fig:transeff}(c). The $k$ exponents are also listed in \tref{tab:exceeding} for different $m$ values. We see that the exponent $k^2$ for the Achlioptas \etal model is approximately twice as large as $k^1$ which is the exponent for the standard percolation model. The exponent $k^m$ keeps increasing with $m$ with a reducing rate. After $m \approx 8$, we find out that the power law behavior starts disappearing since the linear behavior in region (ii) gradually dominate the behavior in region (i) as can be seen in \fref{fig:transeff}(a). The reason for this is the suppressed growth of the largest cluster even just after $p_a^m$. In order to understand the underlying mechanism here, let us note that when there exists many discrete clusters in the lattice with sizes greater than one, adding new bonds joining two discrete clusters is generally favored over adding others that would join the sites of a single cluster. For example, think of a U-shaped cluster with $4$ connected sites exists in the lattice. Let us choose two bond candidates where one of them converts this U-shaped cluster into a unit square with weight $4\times4=16$ and the other connects two discrete clusters of sizes $3$ and $4$ with weight $12$. Obviously, the one with weight $12$ will be occupied even though the total size of the cluster it forms will be greater than that of the unit square (see \Sref{ssec:expperc}). Therefore, the rules we have defined for the growth of lattices support the merging of discrete clusters to form larger ones instead of just `feeding' the present discrete clusters. This fact, of course, leads to an abrupt growth of the largest cluster seen in \fref{fig:transeff}(b) for $m=2$ and $m=4$. Thus, the efficiency increases accordingly since large clusters have more chance to hit both of the opposite sides of the lattice. However, as discrete clusters get more correlated with each other, they are forced to grow in a more specific way from the very beginning, i.e., they grow as homogeneous as possible in size along the lattice as we add the bonds one by one, which eventually results in the suppression of the explosive behavior. Therefore, in order to obtain the most efficient transports, $m$ should be kept at an optimal value and in our case, it is $m=2$ where discrete clusters are `slightly aware' of each other.

In case of an incoherent transport, the excitation can be interpreted as a classical random walker transported with unit efficiency in the limit $t\rightarrow\infty$, provided that it is initially localized on one of the sites from the set $\mathcal{R}$ (see \fref{fig:perclat}) for a given lattice realization. The reason for this is that the walker has enough time to find a correct path towards sink sites within the wrapping cluster. However, in the coherent transport case, the walker may not be able to cross the lattice even if it is initially localized on one of the sites that belong to $\mathcal{R}$. This result originates from the localization effects due to the random scatterings within the disordered structure of the wrapping cluster, namely the Anderson localization \cite{anderson}. In two-dimensions, although the finite size scaling theory suggests that all eigenstates of the system should be exponentially localized independent of disorder strength in thermodynamic limit \cite{abrahams}, it is an ongoing debate whether there are some delocalized states or not due to the different nature of disorder in the percolation model \cite{meir,soukoulis,dillon}. 

In our case, since different $m$ values provide different growth mechanisms, there are structural differences between the clusters they form. For finite lattices, this may lead to different localization effects which directly affect the transport efficiency. We see in \fref{fig:transeff}(d) that coherent transport is slightly inefficient than incoherent transport for all $m$ values even though their behaviors with respect to $p$ are almost completely the same. The differences between their efficiencies are depicted in the inset of \fref{fig:transeff}(d). Although both processes are exposed to the same average disorder, for a given $m$ these slight differences point out the existence of Anderson-type localization in case of coherent transport even for $L=7$. The excitations are obliged to proceed in such disordered paths along the wrapping clusters that they result in destructive interferences, and hence, a decrease in the efficiency. As we expect, the difference is higher for bond fractions where the lattice is highly disordered for each $m$. When $p<0.4$ or $p>0.8$, the difference almost disappears for the lattice transforms into an ordered structure. The cases where the most and the least differences occur are $m=1$ and $m=2$, respectively. This result suggests that, the wrapping clusters formed by choosing $m=2$ are the most scattering ones that eventually prevent the excitations from reaching the sink sites coherently even in the infinite time limit. As we mentioned earlier, this scattering pattern of clusters is highly supported by the $m=2$ case since the selection rule itself favors connecting discrete clusters over placing bonds within a single cluster. In other words, at a certain bond fraction, the total number of bonds for a given wrapping cluster is the least on average for $m=2$. For $m>2$, the efficiency difference starts to reduce since the probability of connecting two discrete clusters gets more equal to the probability of adding a bond to an existing cluster. These results may imply that the amount of correlation between discrete clusters can affect the localization length of eigenstates of coherent transport.

\subsection{\label{ssec:boswc}Localization of eigenstates}

\begin{figure*}[!htbp]
\fontsize{10}{10}\selectfont
\begin{tabular}{cc}
\input{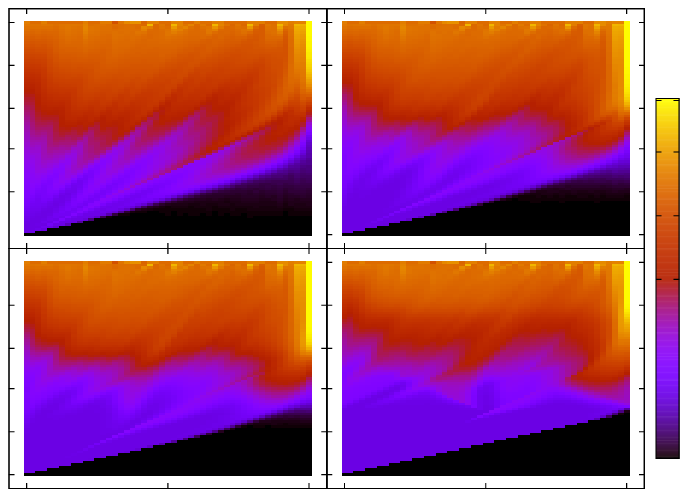}&
\input{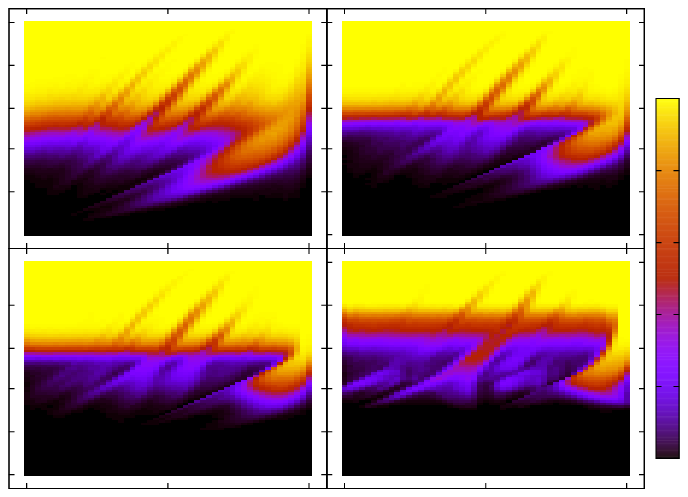}\\
\input{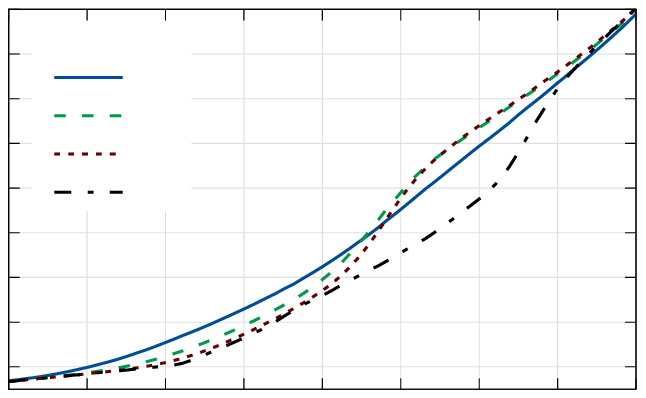}&
\input{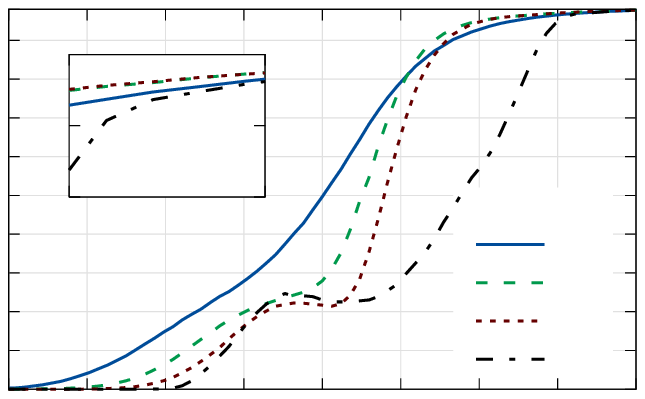}
\end{tabular}
\caption{(a) The average participation ratios $\langle \xi_l(p) \rangle$ of each eigenstate $\ket{\Phi_l^0}$ of the Hamiltonian $H_0$ with respect to the fraction of bonds $p$ in the lattice. $\langle \xi_l(p) \rangle$ is given in the logarithmic scale. (b) The averaged participation ratios $\langle \xi(p) \rangle_\textsubscript{avg}$ over different eigenstates for a given $p$. (c) The contribution probability $\langle \nu_l(p) \rangle$ of each eigenstate $l$ to the transport process. (d) Average number of eigenstates $\gamma(p)$ contributing to the transport. The inset has the same axis labels as the main panel.}
\label{fig:partratio}
\normalsize
\end{figure*}

In order to gain better insight into the localization effects of coherent transport, we can examine each eigenstate of $H_0$ to find out whether they are localized or not. Here we will ignore the trap sites and consider only how disorder affects the eigenstates. Let $\ket{\Phi_l^0(p)}$ be the $l$th eigenstate of $H_0$ for bond fraction $p$. Then, $|\braket{i}{\Phi_l^0(p)}|^2$ gives the probability distribution of the $l$th state over the sites $i\in\mathcal{N}$. The participation ratio provides information about how much a given probability distribution is spread over the lattice and for our case it can be defined as
\begin{equation}
\xi_l(p)=\left( \sum_{i}^N |\braket{i}{\Phi_l^0(p)}|^4 \right)^{-1}.
\label{eq:partratio}
\end{equation}
It estimates the number of sites over which the $l$th eigenstate is distributed, i.e., for $\xi_l=1$ the distribution is localized on a single site whereas $\xi=N$ indicates a homogeneous distribution over $N$ sites. However, we do not acquire any knowledge about the geometrical shape of this distribution, e.g., while a line-shaped cluster with $7$ sites wraps the lattice, a square-shaped cluster with $25$ sites cannot. Moreover, a given distribution may belong to a non-wrapping cluster. Therefore, the participation ratio by itself is not sufficient to decide whether a given eigenstate contributes to transport process or not. For these reasons, let us define a step function $\nu_l (p)$ yielding $1$ if the $l$th eigenstate has at least one nonzero amplitude for each of $\mathcal{S}$ and $\mathcal{S'}$, and yielding $0$ otherwise. The number of source and sink sites involved by the eigenstate is not important since we consider the limit $t \rightarrow \infty$. Thus, the ensemble average $\langle \nu_l (p) \rangle$ becomes the probability of $l$th eigenstate to contribute to the transport process. 

In \fref{fig:partratio}(a), $\langle \xi_l(p) \rangle$ is given for $m=1,2,4$ and $84$. The triangle-shaped black regions represent single-site discrete clusters. For $p=0$, there are $N=49$ of them.  As expected, the number of these clusters is decreased in a linear manner by the addition of more bonds. This transition is most clearly observed in $m=84$ where all single-sited clusters are paired until almost all lattice is covered by two-sited clusters at $p=0.3$. Moreover, different regions are getting separated from each other as we increase $m$. In other words, for a given $p$,
the case where $\langle \xi_l \rangle$ are the most uniformly distributed over $l$ is $m=84.$ This result, of course, arises due to the homogeneous growth of discrete clusters with increasing correlation strength as we have mentioned in \Sref{ssec:treff}. 

We can safely claim that if $\langle \xi_l(p) \rangle < 7$ then $l$th eigenstate is localized and do not contribute to the transportation. It can be seen in Fig. \fref{fig:partratio}(a) that all eigenstates are localized for $p \lesssim 0.2$. For $m=1$ and $m=2$, eigenstates have a chance of contributing to the transport after $p \approx 0.2$ and $p \approx 0.4$, respectively, which also goes along with the results of \fref{fig:transeff}(a) (see $p_a^1$ and $p_a^2$). While $m$ is increasing, we observe that the contributing states are accumulated above $p=0.5$ and support our findings in \fref{fig:transeff}(a) where the transport efficiency starts increasing effectively for $p>0.5$. We note that there are some highly delocalized states near $l=49$. They appear between nearly $p\in[0.75,1]$ for $m=1$ and $m=84$ whereas the same interval becomes $p\in[0.6,1]$ for $m=2$ and $m=4$. For $m=2,4$ and $84$, the sharp transition to these highly delocalized states with increasing $p$ suggests that the wrapping cluster is most likely to reach almost all the sites of the lattice as soon as it is formed. We can further deduce that for $m=1$, the wrapping cluster does not instantly cover the whole lattice when it first appears since the average wrapping probability of $m=1$ is smaller than others. We need to keep adding bonds until $p=0.75$ to make the wrapping cluster cover the lattice. This is one of the main consequences of correlation effects on the transport efficiency.

The average participation ratio of all eigenstates which we define as $\langle \xi(p) \rangle_\textsubscript{avg}=\frac{1}{N}\sum_{l=1}^N \langle \xi_l(p) \rangle $ is shown in \fref{fig:partratio}(b). We see that $\langle \xi \rangle_\textsubscript{avg}$ increases smoothly for $m=1$ whereas it is suppressed up to $p=0.5$ for $m=2,4$ and up to $p=0.7$ for $m=84$, i.e., the greater the value of $m$, the more suppressed is $\langle \xi\rangle_\textsubscript{avg}$. This result also clears up the decrease in efficiency for increasing $m$ values in \fref{fig:transeff}(a). The eigenstates abruptly delocalize following this suppression and get distributed in the lattice over larger regions than the ones in the $m=1$ case. We finally note that the minimum bond fractions satisfying $\left < \xi^{m>1}\right >_\textsubscript{avg} > \left < \xi^{m=1} \right >_\textsubscript{avg}$ are almost the same with $p_b^m$.

The contribution probability $\langle \nu_l (p) \rangle$ depicted in \fref{fig:partratio}(c) is more convenient for understanding whether a given eigenstate contributes to transport or not. The average number of eigenstates contributing to the transport process $\gamma(p)=\sum_l\langle\nu_l(p)\rangle$ is also given in \fref{fig:partratio}(d). We see again that there is almost no contributing eigenstate when $p<0.2$ for all $m$. The only exception is $m=1$ where only a few eigenstates are contributing on average. This early contribution arise from wrapping clusters in the form of lines or strips along the lattice, which we expect to disappear for larger lattices and explains the early rise in the transport efficiency shown in \fref{fig:transeff}(a). On the other hand, nearly all eigenstates are contributing when $p>0.8$ for all $m$. In \fref{fig:partratio}(d), we clearly observe after $p \approx 0.4$ that the average number of contributing eigenstates stays approximately constant, or rather slightly decrease, even though we keep adding more bonds into the lattice. This result is more significant for large $m$ values and it explains the high transport suppression in \fref{fig:transeff}(a). Also, around $p\approx 0.6$ ($p\approx 0.64$), the $\gamma(p)$ for $m=2$ ($m=4$) starts to exceed that of $m=1$ which clears up the high transport efficiencies appear in low bond fractions. If we take into account the similarities between \fref{fig:transeff}(a) and \fref{fig:partratio}(d), it is clear that the transport efficiency is closely related with the number of contributing eigenstates.

\section{\label{sec:con}Conclusion}

We have studied the coherent and incoherent transports between opposite edges of a finite square lattice where the existence of bonds between sites are determined by either standard or explosive percolation models. Since the explosive percolation model provides a disorder source enabling discrete clusters to grow in correlation with each other, we managed to investigate the possible effects of this correlated disorder on the transport efficiency.  We have shown that the least possible correlation is the most contributing to the transport efficiency. For small correlation strengths, we have obtained more efficient transports than that of the standard percolation case. As we increased the correlation strength, the transport efficiency gradually decreased and reduced below that of the standard percolation case. We have demonstrated that the effective starting point and the efficiency of any transport process is directly related to the size of the largest cluster for a given bond fraction.

Moreover, we compared our results with the incoherent transport to see possible localization effects. We have shown that more correlation causes less localization. Therefore, the least correlation provides the most efficient transport despite inducing localization the most. We have explained the possible mechanism behind these localization effects. Lastly, we supported our findings by explicitly examining the average participation ratio and contribution probability of the eigenstates of the system, which allows us to decide whether an eigenstate is localized or delocalized over the lattice.

Depending on our results we conjecture that the localization length of the eigenstates in case of an explosive percolation may be affected by the correlation strength between clusters. A proof of this conjecture, of course, requires careful analysis of the localization properties of eigenstates for larger lattices, which we leave as a topic of further research. Lastly, although there is no strong application of coherent transport on correlated networks yet, our work could provide insight for engineering quantum systems to achieve high efficiency transports by utilizing environmental effects.

\ack
We would like to thank E. \.{I}lker, E. Canay, T. \c{C}a\u{g}lar and O. Benli for helpful discussions.

\appendix
\setcounter{section}{1}
\section*{Appendix: Derivation of the survival probabilities}

We calculate the survival probability $\pi_j$ after starting with the initial state $\ket{\psi_j}$ by summing the transition probabilities over all lattice sites as
\begin{eqnarray}
\pi_j(t)=\sum_{k\in\mathcal{N}}{\pi_{kj}(t)}&=&\sum_{k\in\mathcal{N}}{\bra{\psi_j}e^{iH^{\dagger}t}\ket{k}\bra{k}e^{-iHt}\ket{\psi_j}} \nonumber\\
&=&\bra{\psi_j}e^{iH^{\dagger}t}e^{-iHt}\ket{\psi_j}.
\label{eq:sp}
\end{eqnarray}

\noindent
The Hamiltonian $H=H_0-i\Gamma$ is non-Hermitian and it has $N$ complex eigenvalues $E_l=\epsilon_l-i\gamma_l$ and $E_l^*=\epsilon_l+i\gamma_l$ with respective left $\ket{\Phi_l}$ and right $\bra{\tilde{\Phi}_l}$ eigenstates. These eigenstates can be taken as biorthonormal $\braket{\tilde{\Phi}_l}{\Phi_{l'}}=\delta_{ll'}$ and complete $\sum_{l=1}^{N}\outerp{\Phi_l}{\tilde{\Phi}_l}=I$  \cite{sternheim}. Also, they satisfy $\braket{k}{\Phi_l}^*=\braket{\tilde{\Phi}_l}{k}$.  Therefore, \eref{eq:sp} becomes,
\begin{equation}
\pi_j(t)=\sum_{ll'=1}^N{\braket{\psi_j}{\Phi_l}\bra{\tilde{\Phi}_l}}e^{iH^{\dagger}t}e^{-iHt}\ket{\Phi_{l'}}\braket{\tilde{\Phi}_{l'}}{\psi_j}.
\label{eq:sp2}
\end{equation}

\noindent
By using the following identities,
\begin{eqnarray}
e^{-iHt}\ket{\Phi_{l'}}&=e^{-i\epsilon_{l'}t}e^{-\gamma_{l'}t}\ket{\Phi_{l'}}, \nonumber\\
\bra{\tilde{\Phi}_l}e^{iH^{\dagger}t}&=\bra{\tilde{\Phi}_l}e^{i\epsilon_{l}t}e^{-\gamma_{l}t} \nonumber,
\label{eq:evol}
\end{eqnarray}

\noindent
\eref{eq:sp2} becomes,
\begin{eqnarray}
\pi_j(t)&=\sum_{ll'=1}^N{\braket{\psi_j}{\Phi_l}\braket{\tilde{\Phi}_l}{\Phi_{l'}}\braket{\tilde{\Phi}_{l'}}{\psi_j}}e^{i\epsilon_{l}t}e^{-\gamma_{l}t}e^{-i\epsilon_{l'}t}e^{-\gamma_{l'}t} \nonumber \\ &=\sum_{l=1}^{N}{e^{-2\gamma_l t}\braket{\psi_j}{\Phi_l}\braket{\tilde{\Phi}_l}{\psi_j}}=\sum_{l=1}^{N}{e^{-2\gamma_l t}|\braket{\psi_j}{\Phi_l}|^2}
\label{eq:sp3}
\end{eqnarray}

\noindent
This provides information on how an excitation decays over the lattice in time. In the limit $t \rightarrow \infty$, we expect $\pi_j$ to decay exponentially because of the imaginary parts $\gamma_l$ if the lattice is fully connected. However, when some bonds in the lattice are broken, there exist non imaginary eigenvalues which results in $\lim_{t \rightarrow \infty}\pi_j (t) \ne 0$. Therefore, in \eref{eq:sp3} only the terms with $\gamma_l=0$ may remain and we obtain,
\begin{equation}
\Pi_j=\lim_{t \rightarrow \infty}\pi_j (t) =\sum_{\{l|E_l\in\mathbb{R}\}}|\braket{\psi_j}{\Phi_l}|^2.
\label{eq:limsurprobb}
\end{equation}
\noindent
If we choose an initial state $\ket{\psi_j}=\frac{1}{\sqrt{L}}\sum_{i\in S}\ket{i}$ which is a superposition of $L$ sites from the set $\mathcal{S}$, \eref{eq:limsurprobb} becomes,
\begin{equation}
\Pi_\mathcal{S}=\frac{1}{L}\sum_{\{l|E_l\in\mathbb{R}\}}{\left|\sum_{i\in \mathcal{S}}{\braket{i}{\Phi_l}}\right|^2}.
\label{eq:limsurprobcohini}
\end{equation}
Similarly, we can calculate the survival probability for the incoherent transport as
\begin{equation}
p_j=\sum_{k\in\mathcal{N}}\bra{k}e^{Tt}\ket{\psi_j}=\sum_l \sum_{k\in\mathcal{N}} e^{-\lambda_l t}\braket{k}{\phi_l}\braket{\phi_l}{\psi_j},
\label{eq:incohsurv}
\end{equation}

\noindent
where $-\lambda_l$ ($\lambda_l>0$) and $\ket{\phi_l}$ are the eigenvalues and the corresponding eigenstates of the transfer matrix $T=T_0-\Gamma$. In the $t\rightarrow\infty$ limit, only the terms with $\lambda_l=0$ survive. Therefore, \eref{eq:incohsurv} becomes,
\begin{equation}
P_j=\sum_{\set{l|\lambda_l=0}} \sum_{k\in\mathcal{N}} \braket{k}{\phi_l} \braket{\phi_l}{\psi_j}.
\end{equation}
Then, the initial state $\ket{\psi_j}=\frac{1}{L}\sum_{i\in \mathcal{S}}\ket{i}$  yields,
\begin{equation}
P_\mathcal{S}=\frac{1}{L}\sum_{\set{l|\lambda_l=0}} \sum_{k\in\mathcal{N}} \braket{k}{\phi_l} \left(\sum_{i\in\mathcal{S}} \braket{\phi_l}{i} \right).
\label{eq:limsurprobincohini}
\end{equation}

\section*{References}
\bibliography{bibliography}

\end{document}